# Structured pre-generation elicitation versus single-shot prompting in AI-assisted enterprise decision-making: a randomised online experiment

William Scott-Jackson
Oxford Centre for Impact Research

## Abstract

**Background.** Generative AI speeds, and mostly improves, professional work, but there is concern that users who delegate both the production and the evaluation of an answer may accept weak output and engage less with the underlying reasoning ("cognitive surrender"). Interventions proposed so far, such as unassisted practice or slowing adoption, sit outside the working task. We tested a different approach: an interactive metacognitive scaffolding layer (Cognistance, a prototype developed at the Oxford Centre for Impact Research (OCIR) that asks users to clarify context, choose a strategic direction and explain their reasoning before the AI generates a deliverable.

**Method.** In an online experiment on Prolific, 400 sessions (196 control, 204 treatment) completed an enterprise data-sovereignty crisis task, using either a single prompt or the scaffold. Deliverables were rated on four rubric dimensions (total 4–40) by two LLM raters and one human rater. Participants also completed a seven-item survey and a free-response recall question.

**Results.** Mean composite quality was 32% higher with the scaffold (15.5 vs 20.5; difference 5.0, 95% CI 4.0–6.0; d = 1.02), with the same direction for every rater. Gains were largest for trade-off articulation and strategic coherence and absent for technical specificity. A large part of the aggregate effect reflected rescue of weak prompts: floor-scored (off-task) deliverables fell from 34% to 5%. Among participants whose own prompt already stated the data-localisation problem, the advantage was 21% (d = 1.15). Treatment participants reported greater involvement (d = 0.59) and effort (d = 0.36) and took about 2.4 minutes longer on average (10.46 minutes). Immediate recall scores were higher, which tentatively suggests better retention, but in this limited experiment, was not robust to sensitivity analyses. Self-ratings of quality did not track rated quality in either condition.

**Conclusions.** Structured elicitation before generation improved the rated quality and task relevance of AI-assisted strategy documents at modest cost in time. Delayed retention, error detection and effects in live organisations are the priorities for the next stage of research.

**Keywords:** generative AI; human–AI interaction; metacognitive scaffolding; cognitive offloading; enterprise decision support; output quality; randomised experiment

## 1 Introduction

Generative AI can raise productivity in writing and customer support (Noy & Zhang, 2023; Brynjolfsson et al., 2025), and a field experiment with consultants found large gains for tasks inside the technology's capability frontier and worse performance on a task outside it (Dell'Acqua et al., 2023). In corporate settings a useful document must do more than read well: its priorities need to be clear, its trade-offs defensible and its recommendation understood by the person presenting it. Niederhoffer et al. (2025) describe "workslop", polished AI-assisted material that leaves colleagues to work out what it means and whether it is useful. A related worry is that users who hand over both the drafting and the checking of a recommendation will accept faulty output, learn less from the work and have fewer chances to practise independent skills.

This paper reports a randomised experiment on one design response to that worry. Instead of adding a separate training exercise or limiting use of AI, the intervention changes the working interaction itself: before the AI generates anything, the user answers clarifying questions, selects a strategic direction and explains the choice.

Those inputs both keep the user involved in the reasoning and give the model better information. We compare this with single-shot prompting on the quality of the resulting deliverable, the user's reported involvement and effort, immediate recall of the solution, and the relationship between self-appraisal and rated quality.

### 1.1 From cognitive offloading to cognitive surrender

Cognitive offloading means using external resources to reduce internal processing demands (Risko & Gilbert, 2016); it can free attention for more important judgements. Shaw and Nave (2026) use "cognitive surrender" for the stronger case in which AI output is adopted with minimal scrutiny and displaces the user's own reasoning; in their experiments, accuracy rose when the AI was right and fell when it was wrong. Several lines of work give reasons to take the risk seriously. Lee et al. (2025), in a survey of 319 knowledge workers, found that higher confidence in generative AI was associated with less critical thinking, whereas higher self-confidence was associated with more. Parasuraman and Riley (1997) set out the older automation literature on misuse and disuse. Shukla et al. (2025) analysed UX practitioners' concerns about de-skilling and misplaced responsibility. In a field experiment in high-school mathematics, Bastani et al. (2025) found that unrestricted access to a chatbot improved practice performance but reduced later unaided performance, while a tutor-style design with safeguards largely removed that penalty. The design of the assistance, in other words, appears to matter for what users understand and learn, as well as for what they produce.

### 1.2 Approaches to the problem

Two broad responses can be distinguished. The first, which we call "exercise", maintains capability through deliberate practice, reflection and critical evaluation: retrieval practice has a strong evidence base (Roediger & Karpicke, 2006), and scaffolding and constructive engagement are established routes to active learning (Wood et al., 1976; Chi & Wylie, 2014). Exercise, however, sits outside the working task and depends on users choosing to do extra work. The second, slows or stages AI adoption until governance and skills catch up; the Future of Life Institute's (2023) call for a pause in training the most powerful systems is a prominent example. These proposals address questions well beyond an individual workplace task.

A third option is to build the human contribution into the task. Cognitive forcing functions show that deliberately designed interaction can change reliance on AI (Buçinca et al., 2021; Ghosh et al., 2026), although those interventions act on the review of AI output, and Buçinca et al. found that the benefits were larger for people high in need for cognition and that users rated the most effective designs least favourably. Tankelevitch et al. (2024) argue that generative AI imposes heavy metacognitive demands and that metacognitive support should be built into the systems themselves. The intervention tested here follows that suggestion but acts earlier in the workflow: it elicits the user's decisions before generation and uses them to inform synthesis. Because forcing functions can add friction, we measure effort and time as well as benefit (Sweller, 1988).

### 1.3 Hypotheses

The hypotheses were specified in OCIR design documents before data collection (see Section 2.1). H1 concerns the principal outcome; the others concern secondary or exploratory outcomes.

- **H1.** Mean composite output quality will be higher in the scaffold condition than in the single-shot condition. In the original specification the composite averaged the two LLM raters and the human rater; we report that three-rater mean alongside the two-LLM composite, which covers every session, and use the more conservative of the two as the headline figure.
- **H1a–H1d.** Strategic coherence, technical and regulatory specificity, trade-off articulation and information density will each be higher with the scaffold.
- **H2a.** Self-reported active involvement in shaping the plan (Q4) will be higher with the scaffold. **H2b.** Self-reported mental effort (Q2) will be higher.
- **H3.** Assessor-rated immediate free-response recall of the solution (Q8) will be higher with the scaffold, evaluated separately for each assessor. Delayed retention is outside the scope of this experiment.

- **H4.** The scaled self-appraisal discrepancy, 100 × (Q3/5 − composite quality/40), will be smaller with the scaffold (exploratory).
- **RQ1.** How does recorded task time differ between conditions? Confidence, transparency, reliance and regulatory readiness are examined without directional hypotheses.

# 2 Method

## 2.1 Participants, design, ethics and pre-specification

Participants were recruited through Prolific with filters for management and professional occupations, worldwide. Prolific's own checks were relied on for bot screening, and payment followed Prolific's normal rates (about £2.00 per completed submission). [Add participant demographics if collected.] The export records 400 sessions created between 21 and 25 September 2026, 196 control and 204 treatment. Assignment was random with equal probability at page load. The unit of analysis is the session, identified by the platform field *id*. Prolific identifiers can recur across runs and were not treated as unique participant identifiers; 13 Prolific identifiers appear in more than one session, and Section 3.9 reports a sensitivity analysis.

The Oxford Centre for Impact Research (OCIR) conducted its own ethical review and judged the ethical risk to be very low given the nature of the topic and the participant profile. This was an internal review, not an independent one. The hypotheses, survey instrument and task were specified in OCIR design documents before data collection; they were not entered in a public time-stamped registry. The Holm family, the sensitivity analyses and the prompt-relevance stratification (Section 2.5) were defined after the data were collected and are labelled as such.

## 2.2 Task and conditions

All participants worked on the same fictional scenario. Apex Global is two weeks from commercial launch in Singapore and Malaysia when regulators issue an emergency directive that consumer transaction records, payment details and behavioural analytics must stay on domestic servers; its architecture relies on centralised cloud servers in Frankfurt. The directive is a stipulated condition of the simulation and not a statement of current national law. Participants wrote an initial prompt and received an AI-generated action plan within the Cognistance app, calling, in this case, Gemini 3 1 pro.

In the **control** condition the deliverable was generated from the participant's prompt alone. In the **treatment** condition, after writing the prompt, participants answered three clarification questions (operational priority, infrastructure, primary audience), chose a strategic posture from scaffold cards and wrote a free-text rationale; these inputs informed a synthesis step. The response options are listed in Appendix C. The comparison is between two complete ways of working, not between isolated components, which is the relevant first test of a deployable scaffolding layer. Other functions of Cognistance, such as fact checking and feedback loops, were not applied in this experiment.

## 2.3 Quality ratings

Each deliverable was scored on strategic coherence, technical and regulatory specificity, trade-off articulation, and density or absence of filler, each from 1 to 10 (total 4–40). The rubric prompt is reproduced in Appendix B. Two LLM raters, ChatGPT 6.1 (medium reasoning setting) and Claude Sonnet 5.5, each received the rubric together with only the session *id* and the deliverable. A human rater applied the same rubric to the same inputs. No rater was given the experimental condition, but deliverable style and content may reveal it, so blinding is procedural rather than guaranteed. Two features of the rubric matter for interpretation: it instructs raters to give off-topic deliverables floor scores (total 5–9), and it names a split architecture among examples of a decisive operational posture. Both are discussed in Section 5.

The principal composite is the mean of the two LLM totals (n = 400). The human file covers 398 matched sessions (one human row has no matching session; one control and one treatment session were not rated), and the three-rater mean is reported for those sessions. Every total equals the sum of its four dimensions and all ratings lie within bounds. We report absolute-agreement ICC(2,1) and ICC(2,k), consistency ICC, Krippendorff's interval alpha and Pearson and Spearman correlations (Koo & Li, 2016; Hayes & Krippendorff, 2007). LLM judges can agree well with humans but carry known biases (Zheng et al., 2023), which is a further reason to report each rater separately.

To describe task relevance without relying on the raters' free-text notes, a deliverable is classed as *floor-scored* when at least two of the three raters gave it a total of 9 or less, the top of the rubric's floor range. This classification agrees with the raters' written "off-topic" notes in 399 of 400 sessions.

## 2.4 Survey and free-response recall

Seven post-task items used a 1–5 scale: executive-presentation confidence (Q1), mental effort (Q2), self-rated quality (Q3), active involvement (Q4), perceived transparency (Q5), reliance on AI for core content (Q6) and readiness to defend the plan before regulators (Q7). Wording is in Appendix A. The items are analysed separately and not combined into a scale. Q8 asked: "In your own words, if the company must open on time in 2 weeks, why is setting up a split system usually better than hiring multiple outside local computer firms?

Q8 responses were scored from 0 to 5 by two assessors using a standardised guide: a human (George) and an LLM (ChatGPT 6.1, medium reasoning setting). Both were blind to condition and applied the same rules. The guide scores how accurately and completely a response reflects the scenario and the participant's own generated deliverable. In practice the LLM used scores 0–3 and the human used 0, 3 and 5. Scores exist for 335 sessions: 317 with text and 18 blank responses scored zero. The 65 sessions with no Q8 data are treated as missing, not zero.

## 2.5 Derived measures and prompt-relevance coding

The self-appraisal discrepancy is 100 × (Q3/5 − composite/40). The confidence–recall discrepancy is 100 × (Q1/5 − Q8 score/5), computed for each assessor. These are descriptive indices, not validated measures of metacognitive sensitivity in the sense of Fleming and Lau (2014). Task time is *latency_task_ms* divided by 60,000, from the continue click after briefing to completion of generation; it includes generation waiting time and is not a pure measure of deliberation.

Because every participant wrote an initial prompt before the conditions diverged, a coding of that prompt is unaffected by treatment and can define a legitimate subgroup. A prompt was coded *relevant* if it stated the data-localisation problem (for example, keeping records on in-country servers, moving data out of Frankfurt, or data sovereignty or residency). Coding used a keyword rule (a data term plus a location or jurisdiction term), and we report results both for the rule alone and for the rule plus 16 manual overrides made after reviewing rule-negative prompts and cases where the rule and the raters' floor scores disagreed. The rule was refined after the floor scores were seen, so the coding is not independent of the outcome data; a second blinded coder checked a random 10% sample.

## 2.6 Statistical analysis

Two-sided Welch tests give treatment-minus-control differences with 95% confidence intervals; Cohen's d uses the pooled standard deviation. Percentage differences are the treatment-minus-control difference divided by the control mean, with 95% intervals from 10,000 within-condition bootstrap resamples; because the scale floor is 4, percentages are descriptive ratios of rubric scores and not percentages of business value. Mann–Whitney tests check ordinal and bounded outcomes. A Holm correction (Holm, 1979) is applied to one family of 17 secondary tests: four quality dimensions, seven survey items, two assessor-specific Q8 outcomes, task time, the self-appraisal discrepancy and two confidence–recall discrepancies. Rater-specific estimates, prompt-stratified analyses and other sensitivity checks are not part of the family. Outcome-specific available cases are used and denominators are given; Fisher exact tests compare outcome availability between conditions. Analyses used Python 3.12, pandas, NumPy and SciPy; code is available as described in Section 8.

# 3 Results

## 3.1 Sample and availability

**Table 1** Analysis denominators (sessions)

| Outcome set | Control | Treatment | Total |
|---|---|---|---|
| All sessions; two LLM quality ratings | 196 | 204 | 400 |
| Matched human quality ratings | 195 | 203 | 398 |
| Complete survey (Q1–Q7) and Q8 text | 150 | 167 | 317 |
| Q8 scored (including 18 blank responses) | 160 | 175 | 335 |
| Recorded task time | 183 | 196 | 379 |

Survey availability was 76.5% in control and 81.9% in treatment (Fisher $p = .218$); Q8 scores were available for 81.6% and 85.8% ($p = .280$) and task time for 93.4% and 96.1% ($p = .266$). These tests do not rule out selection effects. Among sessions with complete survey data the composite quality difference was the same as in the full sample (15.41 vs 20.39), so the quality result does not depend on survey completers.

## 3.2 Output quality

The two-LLM composite rose from 15.48 (SD = 5.63) in control to 20.48 (SD = 4.03) in treatment, a difference of 4.99 points (95% CI [4.03, 5.96]; Welch $t(352.2) = 10.17$, $p < .001$; $d = 1.02$, 95% CI [0.82, 1.23]). This is a 32.2% increase on the control mean (bootstrap 95% CI 25.1% to 40.2%). Bootstrap intervals were [4.03, 5.96] for the difference and [0.83, 1.24] for d. H1 is supported. The direction was the same for each rater, but the size differed: ChatGPT 7.19 points ($d = 1.25$), Claude 2.79 ($d = 0.57$) and the human rater 9.15 ($d = 1.35$). The three-rater mean on the 398 matched sessions, the original H1 specification, showed a larger effect (6.39 points, 43.9%), so the two-LLM composite is the more conservative headline.

**Table 2** Output quality: composite and rater-specific estimates (total score, 4–40)

| Outcome | n C/T | Control M (SD) | Treatment M (SD) | Δ [95% CI] | d | % increase [95% CI] | p |
|---|---|---|---|---|---|---|---|
| Two-LLM composite | 196/204 | 15.48 (5.63) | 20.48 (4.03) | +4.99 [4.03, 5.96] | 1.02 | +32.2 [25.1, 40.2] | <.001 |
| ChatGPT | 196/204 | 18.78 (6.91) | 25.98 (4.31) | +7.19 [6.06, 8.33] | 1.25 | +38.3 [30.9, 46.8] | <.001 |
| Claude | 196/204 | 12.19 (4.90) | 14.98 (4.92) | +2.79 [1.83, 3.76] | 0.57 | +22.9 [14.6, 32.3] | <.001 |
| Human rater | 195/203 | 12.67 (7.23) | 21.82 (6.26) | +9.15 [7.81, 10.48] | 1.35 | +72.2 [57.9, 88.7] | <.001 |
| Three-rater mean | 195/203 | 14.55 (5.88) | 20.94 (4.03) | +6.39 [5.39, 7.39] | 1.27 | +43.9 [35.3, 53.4] | <.001 |

Welch tests. Percentage intervals from 10,000 bootstrap resamples.

## 3.3 Quality dimensions

Gains were concentrated in strategic coherence (Δ = 1.94, $d = 1.28$), trade-off articulation (Δ = 1.90, $d = 1.57$) and information density (Δ = 0.89, $d = 0.65$); all survive Holm correction. Technical and regulatory specificity showed a small, non-significant difference (Δ = 0.26, 95% CI [−0.10, 0.62], adjusted $p = .300$). H1a, H1c and H1d are supported; H1b is not.

**Table 3** Quality dimensions (mean of two LLM raters), Holm-adjusted p

| Dimension | n C/T | Control M (SD) | Treatment M (SD) | Δ [95% CI] | d | Holm p |
|---|---|---|---|---|---|---|
| Strategic coherence | 196/204 | 3.60 (1.67) | 5.54 (1.36) | +1.94 [1.64, 2.24] | 1.28 | <.001 |
| Technical and regulatory specificity | 196/204 | 3.15 (1.86) | 3.41 (1.80) | +0.26 [−0.10, 0.62] | 0.14 | .300 |
| Trade-off articulation | 196/204 | 2.89 (1.34) | 4.79 (1.07) | +1.90 [1.66, 2.14] | 1.57 | <.001 |
| Information density | 196/204 | 5.84 (1.73) | 6.73 (0.90) | +0.89 [0.62, 1.16] | 0.65 | <.001 |

### 3.4 Task relevance and prompt quality

The composite effect has two components that single-shot prompting and the scaffold affect differently. First, many control participants wrote prompts that did not state the problem, and the resulting deliverables were off task: 34.2% of control deliverables (67 of 196) were floor-scored against 4.9% of treatment deliverables (10 of 204; Fisher $p < .001$). The same pattern held for each rater (total ≤ 9: ChatGPT 13.8% vs 1.0%; Claude 36.2% vs 18.6%; human 41.3% vs 8.8%). The scaffold supplies the scenario through its questions, so it largely removes this failure.

Second, the scaffold also improved quality among participants who wrote a good prompt. Table 4 splits sessions by the pre-treatment coding of the participant's own prompt. The share of relevant prompts was similar in the two arms (68.4% control, 64.7% treatment). Among participants whose prompt stated the data-localisation problem, treatment still produced a 3.9-point (20.9%) advantage with $d = 1.15$; ChatGPT, Claude and the human rater gave advantages of 5.2, 2.5 and 7.4 points. Among participants whose prompt did not state the problem, 95% of control deliverables were floor-scored against 14% of treatment deliverables, and the advantage was 8.1 points ($d = 2.56$). The effect differed significantly between strata (difference in differences 4.2 points, $z = 6.2$, $p < .001$). Using the keyword rule alone gave the same pattern (relevant prompts: +3.78 points, 20.2%, $d = 1.17$; not relevant: +8.17 points, 90.2%, $d = 2.54$). As a descriptive decomposition, about 59% (bootstrap 95% CI 48% to 71%) of the aggregate difference corresponds to the lower rate of floor-scored deliverables in treatment. That figure conditions on a post-randomisation variable and should not be read causally.

**Table 4** Composite quality by pre-treatment relevance of the participant's own prompt

| Prompt stated the problem? | n C/T | Control M | Treatment M | Δ [95% CI] | d | % increase [95% CI] | Floor-scored C → T |
|---|---|---|---|---|---|---|---|
| Yes | 134/132 | 18.50 | 22.36 | +3.86 [3.05, 4.67] | 1.15 | +20.9 [16.2, 26.0] | 8/134 → 0/132 |
| No | 62/72 | 8.96 | 17.02 | +8.06 [7.00, 9.13] | 2.56 | +90.0 [73.6, 108.1] | 59/62 → 10/72 |
| All sessions | 196/204 | 15.48 | 20.48 | +4.99 [4.03, 5.96] | 1.02 | +32.2 [25.1, 40.2] | 67/196 → 10/204 |

Coding: keyword rule plus 16 manual overrides. Welch intervals for Δ; bootstrap intervals for percentages.

### 3.5 Rater agreement

Absolute-agreement ICC(2,1) was 0.32 for the two LLMs and 0.46 for the three raters; average-measures ICC(2,k) was 0.49 and 0.72. By the guidelines of Koo and Li (2016) these correspond to poor-to-moderate agreement on absolute scores. Interval alpha was 0.08 for the two LLMs and 0.40 for all three, and the two-LLM consistency ICC for the average was 0.80, so the raters agree more on rank order than on level. On the 398 matched sessions the means were 22.46 (ChatGPT), 13.63 (Claude) and 17.34 (human). The composite correlated with the human total at $r = 0.74$ (Spearman 0.67); within condition the correlation was 0.81 in control and 0.42 in treatment, and the mean absolute difference between composite and human was 4.30 points. The supported conclusion is corroborated improvement in rated quality, not a calibrated universal scoring instrument.

### 3.6 Involvement, effort and other self-reports

Active involvement rose from 2.81 to 3.40 ($\Delta = 0.59$, 95% CI [0.37, 0.81], $d = 0.59$, Holm $p < .001$), and mental effort rose by 0.40 points ($d = 0.36$, Holm $p = .021$); H2a and H2b are supported. Confidence and self-rated quality did not differ significantly. Perceived transparency, reliance on AI (lower in treatment) and readiness to defend the plan each had nominal $p < .05$ but did not survive correction, so they are reported as exploratory directions. These are subjective reports; the study did not measure critical-thinking skill directly, and the higher effort rating confirms that the benefit involves additional participation.

**Table 5** Self-reported outcomes (1–5 scale), Holm-adjusted p

| Item | n C/T | Control M (SD) | Treatment M (SD) | Δ [95% CI] | d | Holm p |
|---|---|---|---|---|---|---|
| Q1 Confidence (executive presentation) | 150/167 | 4.11 (0.90) | 4.24 (0.68) | +0.13 [−0.04, 0.31] | 0.17 | .300 |
| Q2 Mental effort | 150/167 | 3.48 (1.27) | 3.88 (0.96) | +0.40 [0.15, 0.65] | 0.36 | .021 |
| Q3 Self-rated quality | 150/167 | 4.05 (0.90) | 4.22 (0.70) | +0.17 [−0.01, 0.35] | 0.21 | .261 |
| Q4 Active involvement | 150/167 | 2.81 (1.05) | 3.40 (0.94) | +0.59 [0.37, 0.81] | 0.59 | <.001 |
| Q5 Perceived transparency | 150/167 | 3.94 (0.91) | 4.16 (0.78) | +0.22 [0.03, 0.40] | 0.26 | .168 |
| Q6 Reliance on AI | 150/167 | 4.15 (1.03) | 3.89 (1.02) | −0.26 [−0.49, −0.03] | -0.26 | .168 |
| Q7 Readiness to defend | 150/167 | 3.78 (0.95) | 4.03 (0.93) | +0.25 [0.04, 0.46] | 0.27 | .152 |

## 3.7 Immediate free-response recall

Q8 scores were higher in treatment for both assessors: the LLM assessor 0.83 to 1.20 (Δ = 0.38, 95% CI [0.12, 0.63], d = 0.31) and the human assessor 1.64 to 2.19 (Δ = 0.56, 95% CI [0.18, 0.94], d = 0.31); both have Holm-adjusted p = .043. H3 is supported at this level, and we regard it as a tentative signal. Assessor agreement was modest (r = 0.41, Spearman 0.40, ICC(2,1) = 0.32, alpha = 0.27), and the human assessor scored 0.91 points higher on average on a coarser scale.

The effect was sensitive to how the sample was defined (Table 6). Excluding floor-scored deliverables, the differences fell to 0.16 points for the LLM assessor (95% CI [−0.16, 0.47]) and 0.33 for the human assessor ([−0.12, 0.78]), neither significant. Within the stratum of participants whose own prompt stated the problem the human assessor still found a difference (0.59, p = .018) and the LLM assessor did not (0.22, p = .20). Two features of the measure help explain this. Q8 asks about a split system, which most treatment participants had chosen in the scaffold and which appears in nearly all treatment deliverables, and answers are scored against the participant's own deliverable, which differs by condition. Participants in the two conditions mentioned a split, local or in-country arrangement in their answers at the same rate (52.0% control, 52.7% treatment). The LLM assessor also flagged 70 control records and 18 treatment records for review, typically because the deliverable concerned another task. Q8 therefore shows that treatment responses aligned better with the deliverables participants received; it does not show better retention independent of that deliverable, and delayed retention was not measured.

**Table 6** Immediate free-response recall (Q8, 0–5 guide) by assessor

| Assessor | n C/T | Control M (SD) | Treatment M (SD) | Δ [95% CI] | d | Holm p | Δ excluding floor-scored [95% CI] |
|---|---|---|---|---|---|---|---|
| LLM (ChatGPT 6.1) | 160/175 | 0.83 (1.16) | 1.20 (1.23) | +0.38 [0.12, 0.63] | 0.31 | .043 | +0.16 [−0.16, 0.47] |
| Human | 160/175 | 1.64 (1.77) | 2.19 (1.78) | +0.56 [0.18, 0.94] | 0.31 | .043 | +0.33 [−0.12, 0.78] |

## 3.8 Self-appraisal and rated quality

The self-appraisal discrepancy was smaller in treatment (42.54 vs 33.45 scaled points; Δ = −9.09, 95% CI [−13.55, −4.63], d = -0.46, Holm p = .001). This arose because rated quality rose while self-ratings changed little, and not because users judged their own work more accurately: self-rated quality correlated with composite quality at only r = .07 overall (.03 in control, .02 in treatment). H4 is therefore supported only as a descriptive result about the gap between two rating sources. The confidence–recall discrepancy was not significantly different after correction for either assessor (LLM Δ = −4.89, Holm p = .30; human Δ = −8.40, Holm p = .24).

## 3.9 Time and robustness

Mean task time was 10.52 minutes in control and 12.89 in treatment (Δ = 2.37 minutes, 95% CI [0.99, 3.75], d = 0.35, Holm p = .009); medians were 9.13 and 11.66 minutes (about 28% longer; Mann–Whitney p < .001).

The 21 sessions that never completed (13 control, 8 treatment) have no time record. The quality gain per added minute (2.11 composite points) is descriptive and is not a productivity or financial estimate. Restricting the sample to the first-created session for each Prolific identifier (387 sessions) gave composite means of 15.59 and 20.61 ($\Delta = 5.02$, 95% CI [4.04, 6.00]), so repeated identifiers do not drive the result, although individual participants cannot be uniquely identified.

## 4 Discussion

Asking users to clarify, choose and justify before generation produced higher-rated deliverables, and every rater saw the improvement. The gain came by two routes. For participants whose own prompt did not state the problem, the scaffold turned mostly off-task output into usable plans. For participants who did state it, the scaffold still raised quality by about a fifth, a moderate-to-large effect, and the human rater saw a larger one than either LLM. The first route is a practical result in its own right: a structured front end is a guard against the irrelevant output that single-shot prompting produced in a third of control sessions. The second suggests that the benefit is not only about supplying missing context.

The pattern across dimensions is informative. Coherence, trade-off articulation and density improved, while technical and regulatory specificity did not. The scaffold elicits a decision and a justification; it does not supply statute-level evidence, so specialist knowledge and checking remain necessary in a corporate setting. The qualities that improved are those that help a colleague understand and challenge a recommendation, which bears on the workslop problem described by Niederhoffer et al. (2025). Whether better-rated documents reduce review time or improve decisions needs testing in live workplaces.

Participants in the scaffold condition reported feeling more actively involved and putting in more effort, at a cost of about two and a half minutes. That is consistent with the aim of keeping the human in the reasoning, and with Tankelevitch et al.'s (2024) argument for metacognitive support inside the tool. The effect on involvement is moderate; whether it carries over to the capability the cognitive-surrender literature is concerned with, namely scrutiny of faulty output, is a separate question that this design does not test (Shaw & Nave, 2026). Higher immediate recall scores in treatment are a suggestive signal that the interaction may help users grasp the solution. The sensitivity analyses show that the signal is not yet robust, and a delayed, deliverable-independent test is the natural next step, particularly given evidence that assisted performance and later unaided performance can diverge (Bastani et al., 2025). Self-ratings of quality tracked rated quality poorly in both conditions, in line with the disconnect between performance and metacognition reported by Fernandes et al. (2026), so the scaffold improved output without, on this evidence, improving users' ability to judge it.

## 5 Limitations and priorities for further research

The study was designed as a first test of a deployable scaffolding layer, and each of its boundaries points to a specific next study.

- **Which component matters.** The comparison is between two complete workflows. An arm that supplies the same structured inputs without interaction, and arms that separate choosing from explaining, would show whether involvement itself adds value beyond better information.
- **Comparators and the cognitive-surrender claim.** There is no unaided arm, no AI-alone arm (Vaccaro et al., 2024) and no trials with deliberately faulty AI output. Testing whether scaffolding reduces uncritical acceptance of errors requires such trials (Shaw & Nave, 2026).
- **Rubric and raters.** The rubric is domain-specific and was written by the study team. It names a split architecture as an example of a decisive posture, which overlaps with an option in the scaffold, and it instructs raters to floor-score off-topic work, which contributes mechanically to the off-task difference. The LLM raters differ markedly in severity and agree only moderately on absolute scores; one human rater scored quality; and one Q8 assessor and one quality rater come from the same model family. An independent expert panel, a rubric that is neutral between architectures, and rating of style-matched deliverables would strengthen construct validity (Zheng et al., 2023).

- **Retention.** Q8 was an immediate free-response item, phrased around a split system and scored against each participant's own deliverable. A delayed test with neutral items and an unaided application task would show whether the suggested retention benefit is real and lasting; related work is beginning to examine how well users remember AI-assisted work (Zindulka et al., 2026).
- **Population and setting.** Participants were Prolific selected professionals completing a short simulated task. Field trials in organisations, measuring revision effort, readiness to act and decision outcomes, are the most valuable next step.
- **Diversity of output.** Most treatment participants chose the same posture (165 of 204 selected the split-system option). AI assistance can raise individual output quality while narrowing the range of ideas (Doshi & Hauser, 2024), so future studies should measure diversity as well as quality.

## 6 Conclusion

In an enterprise simulation, building structured user decisions into the AI workflow produced deliverables that were rated about a third higher on average than those from single-shot prompting, removed most off-task output, and increased reported involvement at a cost of about two and a half minutes. The advantage remained at about one fifth among participants who wrote a good prompt themselves. The results support corporate trials of the approach, with delayed retention, error detection and independent replication as the priorities for evaluating whether it also protects understanding and independent skill over time.

## 7 Declarations

### 7.1 Use of generative AI

Generative AI tools assisted in preparing this work. Gemini was used for an initial manuscript draft and ChatGPT for an earlier revision. Claude Sonnet 5.5 (Anthropic) was used to re-analyse the supplied data, check references and redraft this version; all statistics in this version were recomputed from the source files with executable code. AI was also part of the experimental task (as the generator of deliverables) and of the measurement (ChatGPT 6.1 and Claude Sonnet 5.5 as quality raters, and ChatGPT 6.1 as a Q8 assessor); these research uses are distinct from manuscript assistance. AI tools are not authors. The author has reviewed and approved the manuscript, analyses and references and takes responsibility for the content.

### 7.2 Ethics, funding and competing interests

Ethical review was conducted internally by OCIR, which assessed the risk as very low. Participants took part through Prolific under its terms covering privacy, data protection and ethical treatment. The research was funded internally by OCIR. **Competing interests:** the tool evaluated here, Cognistance, was invented by OCIR, which may develop it for commercial application; the author is affiliated with OCIR. OCIR staff designed the scenario, rubric and scaffold, and reviewed them using human and multiple LLM checks. Neither the author nor OCIR has a financial relationship with any AI developer or software company. Independent replication is welcomed.

## 8 Data and code availability

De-identified session data, the rater and assessor prompts and scoring guide, the scaffold specification, the prompt-relevance coding and the analysis code are held at www.oxfordcentreforimpactresearch.com [insert exact URL]. Participant-written free text is not released if it could identify individuals.

## Appendix A Survey instruments

Q1 Confidence. How confident would you feel presenting the strategy produced today to an executive board?

Q2 Effort. How much mental focus and cognitive effort did this task require?

Q3 Quality. How highly would you rate the quality of your action plan?

Q4 Involvement. How actively involved did you feel in shaping the strategic direction of the final plan? (1 = Completely passive … 5 = Fully in control)

Q5 Transparency. How transparent did the AI's reasoning feel while it generated the plan?

Q6 Reliance. How much did you rely on the AI to produce the core strategic content?

Q7 Readiness. How ready do you feel to defend this plan if questioned by regulators?

Q8 Recall (open response, at least 20 characters). In your own words, if the company must open on time in 2 weeks, why is setting up a split system usually better than hiring multiple outside local computer firms?

Q1–Q7 are scored 1–5 (1 = low, 5 = high).

## Appendix B Quality-rating prompt (verbatim)

You are serving as an independent, double-blind expert evaluator in a computational social science randomized controlled trial (RCT). EVALUATION CONTEXT: The scenario evaluates an urgent enterprise data sovereignty crisis in Southeast Asia (Apex Global). An emergency regulatory directive in Singapore and Malaysia bans the cross-border transfer of raw customer PII, payments, and behavioral analytics to the central cloud datacenter in Frankfurt, Germany, with only two weeks before commercial launch. Use the field named 'id' as the unique identifier for each response.

EVALUATION RUBRIC (score each dimension strictly from 1 to 10): (1) Strategic Coherence: decisive, actionable operational posture (e.g., split architecture, selective edge colocation, deliberate build delay) vs. non-committal corporate generalities and process audits. (2) Technical & Regulatory Specificity: concrete handling of Singapore PDPA and Malaysia PDPA 2010, Frankfurt cloud egress controls, tokenization/PII redaction, and network latency constraints. (3) Trade-off Articulation: explicit balancing of architectural complexity, vendor fragmentation, technical debt, and launch timeline friction vs. frictionless, unrealistic "magic-bullet" assertions. (4) Density & Absence of Filler: 10 = dense, high-signal executive memo format with zero boilerplate; 1 = generic, repetitive AI filler, conversational platitudes, or off-topic responses.

SCORING INSTRUCTIONS: Off-topic deliverables (e.g., social media marketing, personal investment, café business plans) MUST receive the absolute minimum floor scores (Coherence: 1–2, Specificity: 1, Tradeoff: 1, Density: 2–3, Total: 5–9). Total Score = coherence + specificity + tradeoff + density (scale 4 to 40). Be strict regarding implementation realism and statutory compliance. INPUT: participant deliverable table containing id and generated_deliverable. OUTPUT: only a raw CSV block with the header id, coherence, specificity, tradeoff, density, total_score, justification_notes.

## Appendix C Treatment scaffold options and choices

Treatment participants (n = 204) selected from the following options. **Operational priority:** ensure 100% regulatory compliance regardless of delays (159); launch on planned date at all costs (33); minimise infrastructure capital expense (12). **Infrastructure:** split system keeping private customer data isolated on local servers while sending non-sensitive tasks to the cloud (165); contract regional third-party colocation facilities (39). **Primary audience:** CEO – market share, brand risk, timeline (135); CTO – data pipelines, system latency, technical debt (69). **Strategic posture (scaffold cards):** split architecture (125); compliance first (55); speed first (24). Participants then wrote a free-text rationale. The Q8 scoring guide and the full scaffold specification are in the data and code repository (Section 8).